\documentclass[runningheads]{llncs}
\usepackage{graphicx}
\usepackage{bbding}
\usepackage{amsmath}
\usepackage{cite}
\begin{document}
\title{A Multi-Stage Framework for the 2022 Multi-Structure Segmentation for Renal Cancer Treatment}
%
%
\author{Yusheng Liu \and
Zhongchen Zhao\and
Lisheng Wang\inst{(}\Envelope\inst{)}}
\authorrunning{F. Author et al.}
%
\institute{Institute of Image Processing and Pattern Recognition, Department of Automation,
Shanghai Jiao Tong University, Shanghai 200240, People’s Republic of China\\
\email{\{13193491346,lswang\}@sjtu.edu.cn}}
\maketitle              
\begin{abstract}
Three-dimensional (3D) kidney parsing on computed tomography angiography (CTA) images is of great clinical significance. Automatic segmentation of kidney, renal tumor, renal vein and renal artery benefits a lot on surgery-based renal cancer treatment. In this paper, we propose a new nnhra-unet network, and use a multi-stage framework which is based on it to segment the multi-structure of kidney and participate in the KiPA2022 challenge.

\end{abstract}
\section{Introduction}
Kidney cancer accounts for nearly 3\% of all cancers\cite{2018Global}, for which the reference treatment is surgery-based. Among all the procedures of the renal cancer treatment, three-dimensional (3D) kidney parsing on computed tomography angiography (CTA) images is occupied a crucial position, targeting the segmentation of kidney\cite{2019Coarse}, renal tumors\cite{2021Leveraging}, arteries as well as veins \cite{HE2021}.
\par
With the help of automatic kidney segmentation, clinicians could observe the lesion regions and draw up the plan more easily and accurately, putting more emphasis on intraoperative treatment. Laparoscopic partial nephrectomy (LPN)\cite{SHAO2011849}, as a popular method of minimally invasive surgery\cite{2022Organ}, focuses on clamping branches of the renal artery, which also depends on the accurate renal vascular segmentation\cite{SHAO20121001}. Based on the great clinical significance, KiPA2022 aims to accelerate the development of reliable, valid, and reproducible methods to reach this need, so as to promote surgery-based renal cancer treatment.

\section{Methods}
Deep convolutional network has been proven to be a reliable tool for semantic segmentation, especially in the medical scenario. The appearance of more and more efficient algorithms, including 3DU-Net\cite{iek20163D}, nnU-Net and so forth, has actually provided abundant idea with us. In this paper, we develop a multi-stage strategy to effectively and efficiently segment the kidney, the kidney tumor, the renal vein and the renal artery, as shown Fig.1. This framework is composed of three stages of segmentation with a carefully designed nnhra-unet. We also use two loss functions highly related with hard-region-adaption and design a systematic pipeline including several process according to imaging itself to make the segmentation more accurate. To be specific, our algorithm contains four steps:
\subsubsection{Coarse Segmentation.} We use two distinct nnhra-unet to get the coarse segmentations respectively, between which the loss funtions we choose are different. One is for the kidney and tumor part, and another is for the vessel part. In terms of the coarse segmentation I, we use the self-adapting crop skill to get the bounding box (bbox) containing the tumor region-of-interest and expand the bbox from 1.0 to 1.5 times according to the case. As we shrink the image size, the network could emphasize on the only tumor part, which leads to the greater segmentation results. When it comes to the coarse segmentation II, the proposal aims at the vessel segmentation and will be told in Fine Vein \& Artery Segmentation in detail.
\subsubsection{Fine Kidney Segmentation.}We obtain the predicted mask of kidney from the coarse segmentation I by four-classification nnhra-unet as the fine kidney segmentation.
\subsubsection{Fine Tumor Segmentation.}We first combine the coarse tumor predictions from the coarse segmentation I \& II through voting. With the tumor ROI and the coarse tumor segmentation, we segment the tumor by a single-classification nnhra-unet and use a cyst filter module to restrain the FP structure in order to get the fine tumor segmentation.
\subsubsection{Fine Vein \& Artery Segmentation.}Considering the complex arteriovenous separation, we obtain the renal vein and artery segmentation simultaneously from the coarse segmentation I \& II via ensemble learning strategy.
\begin{figure}
\includegraphics[width=\textwidth]{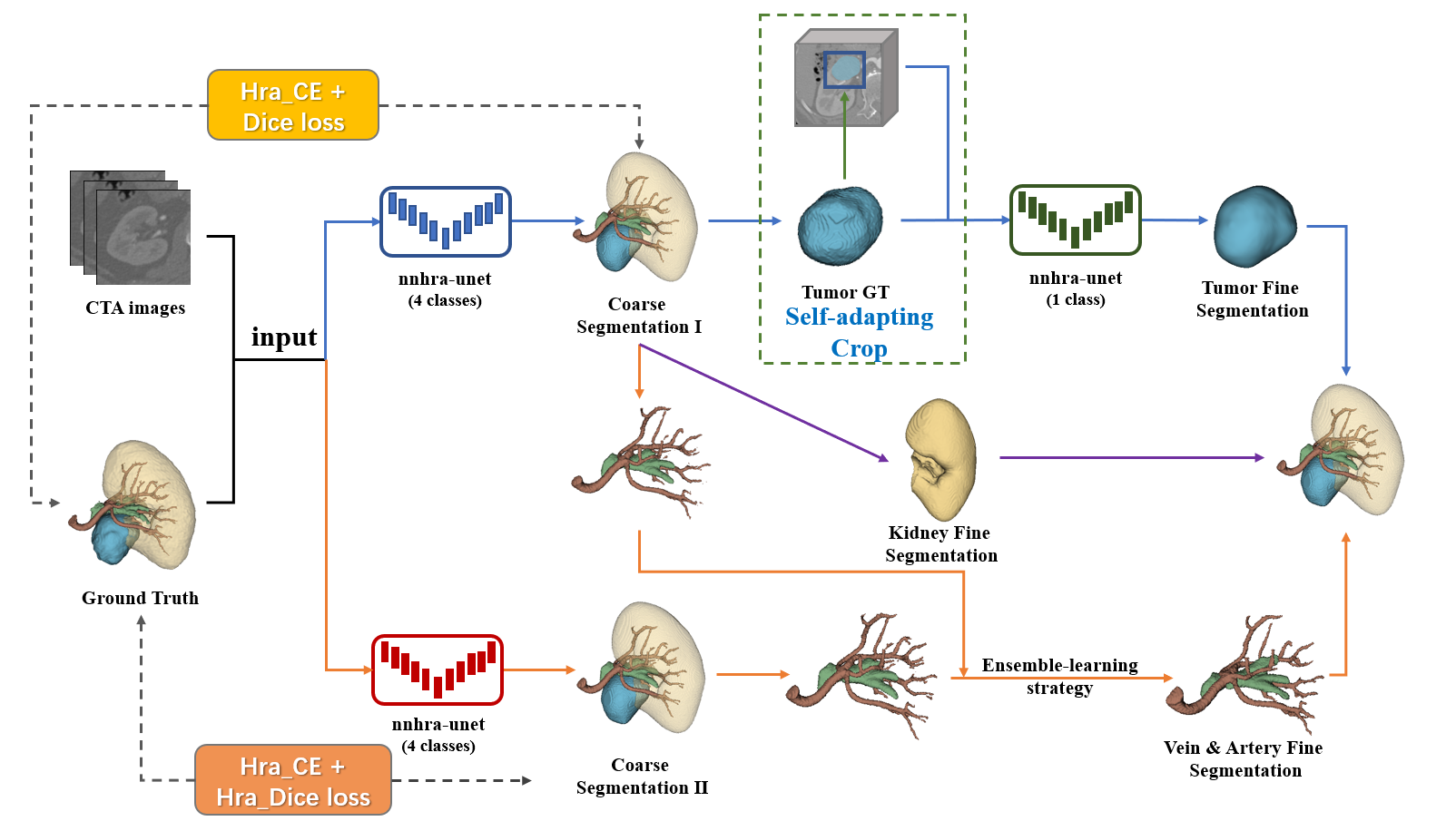}
\caption{An overview of our multi-stage segmentation framework.} \label{fig1}
\end{figure}
 
\subsection{Preprocessing}
The baseline method includes the following preprocessing steps:
\begin{itemize}
\item[$\bullet$]Resampling. All the images are resampled to the same target spacing [0.63281, 0.63281, 0.63281] by using the linear spline interpolation so that the resolution of the z-axis is similar to that of the x-/y-axis. The patch size of the coarse and fine input are [160, 128, 112] and [80, 80, 80], respectively. 
\vspace{0.5cm}
\item[$\bullet$]Intensity Normalization. A z-score normalization is applied based on the mean and standard deviation of the intensity values from the foreground of the whole dataset.
\vspace{0.5cm}
\item[$\bullet$]Data Augmentation. After resampling and normalizing the dataset, we use several data augmentation tricks, including rotation, scaling, mirroring, etc. in the process of training.
\end{itemize}
\subsection{Proposed Method}
The proposed nnhra-unet consists of two major parts: a nnU-Net\cite{0nnU} as the baseline, and the hard-region-adaption loss functions\cite{HE2020} .
\setcounter{secnumdepth}{4}
\subsubsection{Hard-Region-Adaption Loss}
\ 
\newline
After analyzing the dataset, we find that both renal artery and vein account for only a little precent of the whole image, resulting in the class imbalance. Considering that, we decide to use the hard region adaptation (HRA) loss function\cite{HE2020}, which will penalize the hard region of the vessels structure in the image, such as vascular boundary, small ends, etc.\cite{2020clDice} and put more emphasis on the boundary of the kidney structure as well.
\par
In this paper, we apply this inter-image sampling strategy to both cross-entropy and dice loss. A detail description of the modified loss functions is as follows.
\begin{itemize}
\item[$\bullet$]The HRA-CE loss function is defined as the special cross-entropy loss, covering the hard-to-segment area instead of all regions, as shown below.
\begin{center}
\begin{equation}
L_{HRA-CE} = -\frac{1}{N}\sum_{c = 1}^C\sum_{n = 1}^N I(y_{n,c},\hat{y}_{n,c})y_{n,c}\log{\hat{y}_{n,c}}
\end{equation}
\end{center}
where $y_{n,c}$ is the ground truth of the cth class and the nth pixel, while $\hat{y}_{n,c}$ is the predicted value. $I(y_{n,c},\hat{y}_{n,c})$ is a region-choosing strategy that thresholds for the hard-to-segment mask through the judgement of the L1 distance between the target value and predicted value. If $|y_{n,c}–\hat{y}_{n,c}| < T$, $I(y_{n,c},\hat{y}_{n,c})$ equals 1, othervise $I(y_{n,c},\hat{y}_{n,c})$ equals 0. When T is set as 0, the HRA-CE loss is the same as CE loss.\\
\vspace{0.5cm}
\item[$\bullet$]The HRA-Dice loss function is defined the same as the HRA-CE loss, as shown below.
\begin{center}
\begin{equation}
L_{HRA-Dice} = -\frac{1}{N}\sum_{c = 1}^C\sum_{n = 1}^N I(y_{n,c},\hat{y}_{n,c})\frac{2y_{n,c}\hat{y}_{n,c}}{y_{n,c}^2\hat{y}_{n,c}^2}
\end{equation}
\end{center}
\end{itemize}
In our training process, we combine the HRA-CE loss with Dice loss in the coarse segmentation I to get the more accurate segmentation of renal structure and prevent the over-segmentation from paying excessive attention to the marginal context information by the dice loss with HRA part. Conversely, to guide the model to learn more minor details of the vessel structure,we suppose to use the combination with HRA-CE loss and HRA-Dice loss in the coarse segmentation II.
\subsubsection{Kidney Segmentation}
\ 
\newline 
After experimental training, we find that the accuracy of kidney is better than the accuracy of the tumor and vessel because of its appropriate receptive field. So we decide to first directly segment the kidney mask by the nnhra-unet in the coarse segmentation I \& II and then segment the tumor from the kidney mask. Importantly, we regard the predicted kidney mask from segmentation I as the fine kidney segmentation.
\subsubsection{Tumor Segmentation}
\ 
\newline
With the predicted tumor mask from the coarse segmentation I \& II, we get the tumor ROI from the CTA image for each case via self-adapting crop strategy. To be more specific, we shrink the ROI into tumor domain with accordance to connected domain which is larger than 2000 pixels. Then, we feed the cropped image as well as predicted mask containing one single potential tumor part into the nnhra-unet.
\subsubsection{Renal Vein \& Artery Segmentation}
\ 
\newline
For the renal vein part, we suppose to integrate the vein mask from the coarse segmentation I \& II via ensemble learning method at a ratio of 4:6. For the renal artery part, we adopt 3D-based average pooling with the stride of $3\times3\times3$ to extract the minor feature of the artery mask from the coarse segmentation II, and combine the thin structure of the artery with the main part from the coarse segmentation I.

\subsubsection{Postprocessing}
\ 
\newline
After model inference, we put the predicted results into a systematic process. First, we keep the max connected domain of the kidney. When it comes to vein and artery, we remove the connected domain whose center point distance from the center point of the max connected domain is longer than 100 and 92 respectively, and we also remove the domain that the CT value is higher than 2200 for the artery because of the radiographic fact that the domain with high CT value is the probably the False Postive part (FP) where contains other kind of liquid than blood. Regarding to the tumor part, we first propose the cyst filter to restrain the FP as cyst and tumor have difference on the CT value, and then remove the connected domain which are smaller than 100. Finally, we combine the fine kidney, tumor, vein and artery mask together as the fine segmentation.
\subsubsection{Implementation Details}
\ 
\newline
During the training, the batch size is set as 2 and the batch normalization (BN) is our choice. Moreover, we use stochastic gradient descent (SGD) as the optimizer. The initial learning rate is set to be 0.01 and the total training epochs are 300. The patch size is [160, 128, 112] in both the coarse segmentation I \& II while [80,80,80] in the fine tumor segmentation stage. We use the 5-fold cross-validation in nnhra-unet to get greater model performance, and other hyper-parameters mainly follow the baseline nnU-Net as default. We implement our network with PyTorch based on a single NVIDIA GeForce RTX 3090 GPU with 24 GB memory.

\section{Dataset and Evaluation Metrics}
\subsection{Dataset}
\begin{itemize}
\item[$\bullet$]Description of the dataset used:\\
The dataset from KiPA2022 is annotated by mature algorithms and adjusted by experts. Instead of double renal region, the images only contain right or left part of the kidney structure and both the kidney and tumor labels are expanded to a maximum of 32 pixels to crop the ROI regions. For more detail information, please refer to the challenge website.
\vspace{0.5cm}
\item[$\bullet$]Details of training and validation data:\\
The total number of cases is 130. 70/30/30 of the original images are selected in Training/Opening test/Closing test phase respectively.
\vspace{0.5cm}
\item[$\bullet$]In our training process, the dataset (70 cases) is randomly divided into training and validation set at a ratio of 4:1. A 5-fold cross validation set is generated on the above partition.
\end{itemize}
\subsection{Evaluation Metrics}
\begin{itemize}
\item[$\bullet$]Dice Similarity Coefficient (DSC)
\vspace{0.5cm}
\item[$\bullet$]Hausdorff Distance (HD)
\vspace{0.5cm}
\item[$\bullet$]Average Hausdorff Distance (AVD)
\end{itemize}

\bibliographystyle{abbrv}
\bibliography{1.bib}  

\begin{thebibliography}{10}

\bibitem{2018Global}
F.~Bray, J.~Ferlay, I.~Soerjomataram, R.~L. Siegel, and A.~Jemal.
\newblock Global cancer statistics 2018: Globocan estimates of incidence and
  mortality worldwide for 36 cancers in 185 countries: Global cancer statistics
  2018.
\newblock {\em CA A Cancer Journal for Clinicians}, 68(suppl 8), 2018.

\bibitem{HE2020}
Y.~He, G.~Yang, J.~Yang, Y.~Chen, Y.~Kong, J.~Wu, L.~Tang, X.~Zhu, J.-L.
  Dillenseger, P.~Shao, S.~Zhang, H.~Shu, J.-L. Coatrieux, and S.~Li.
\newblock Dense biased networks with deep priori anatomy and hard region
  adaptation: Semi-supervised learning for fine renal artery segmentation.
\newblock {\em Medical Image Analysis}, 63:101722, 2020.

\bibitem{HE2021}
Y.~He, G.~Yang, J.~Yang, R.~Ge, Y.~Kong, X.~Zhu, S.~Zhang, P.~Shao, H.~Shu,
  J.-L. Dillenseger, J.-L. Coatrieux, and S.~Li.
\newblock Meta grayscale adaptive network for 3d integrated renal structures
  segmentation.
\newblock {\em Medical Image Analysis}, 71:102055, 2021.

\bibitem{iek20163D}
z.~iek, A.~Abdulkadir, S.~S. Lienkamp, T.~Brox, and O.~Ronneberger.
\newblock 3d u-net: Learning dense volumetric segmentation from sparse
  annotation.
\newblock In {\em Springer, Cham}, 2016.

\bibitem{0nnU}
F.~Isensee, P.~F. Jaeger, S.~A.~A. Kohl, J.~Petersen, and K.~H. Maier-Hein.
\newblock nnu-net: a self-configuring method for deep learning-based biomedical
  image segmentation.
\newblock {\em Nature Methods}.

\bibitem{2019Coarse}
S.~Liu.
\newblock Coarse to fine framework for kidney tumor segmentation.
\newblock In {\em 2019 Kidney Tumor Segmentation Challenge: KiTS19}, 2019.

\bibitem{2021Leveraging}
C.~B. Lund and D.~Van.
\newblock Leveraging clinical characteristics for improved deep learning-based
  kidney tumor segmentation on ct.
\newblock 2021.

\bibitem{2022Organ}
A.~Mcdonald-Bowyer, S.~Dietsch, E.~Dimitrakakis, J.~M. Coote, L.~Lindenroth,
  D.~Stoyanov, and A.~Stilli.
\newblock Organ shape sensing using pneumatically attachable flexible rails in
  robotic-assisted laparoscopic surgery.
\newblock 2022.

\bibitem{SHAO2011849}
P.~Shao, C.~Qin, C.~Yin, X.~Meng, X.~Ju, J.~Li, Q.~Lv, W.~Zhang, and Z.~Xu.
\newblock Laparoscopic partial nephrectomy with segmental renal artery
  clamping: Technique and clinical outcomes.
\newblock {\em European Urology}, 59(5):849--855, 2011.

\bibitem{SHAO20121001}
P.~Shao, L.~Tang, P.~Li, Y.~Xu, C.~Qin, Q.~Cao, X.~Ju, X.~Meng, Q.~Lv, J.~Li,
  W.~Zhang, and C.~Yin.
\newblock Precise segmental renal artery clamping under the guidance of
  dual-source computed tomography angiography during laparoscopic partial
  nephrectomy.
\newblock {\em European Urology}, 62(6):1001--1008, 2012.

\bibitem{2020clDice}
S.~Shit, J.~C. Paetzold, A.~Sekuboyina, A.~Zhylka, and B.~H. Menze.
\newblock cldice - a topology-preserving loss function for tubular structure
  segmentation.
\newblock 2020.

\end{thebibliography}
\end{document}